\documentclass[twocolumn,amsmath,amssymb,floatfix,prd]{revtex4}
\usepackage{bm,epsfig}
\usepackage{amsmath}
\usepackage{epsf}

\newcommand{\beq}{\begin{equation}}

\newcommand{\eeq}{\end{equation}}
\newcommand{\beqa}{\begin{eqnarray}}
\newcommand{\eeqa}{\end{eqnarray}}

\def\simgt{\gtrsim}

\newcommand{\bmax}{b_{\rm max}}

\newcommand{\np}{{n_{\rm p}}}
\newcommand{\nm}{n_{\rm m}}
\newcommand{\vecdelta}{{\boldsymbol \delta}}
\newcommand{\veclambda}{{\boldsymbol \lambda}}
\newcommand{\vecSigma}{{\boldsymbol \Sigma}}

\begin{document}

\title{Likelihood Methods for Cluster Dark Energy Surveys}
\author{Wayne Hu$^{1}$ and J.D. Cohn$^{2}$ }
\affiliation{
$^{1}$Kavli Institute for Cosmological Physics, Department of Astronomy and Astrophysics,
and Enrico Fermi Institute,
University of Chicago, Chicago IL 60637 \\
$^{2}$Space Sciences Laboratory and Theoretical Astrophysics Center, 
University of California, Berkeley CA 94720
}

\begin{abstract}
\baselineskip 11pt 
Galaxy cluster counts at high redshift, binned into spatial pixels 
and binned into ranges in an observable proxy for mass, contain a wealth 
of information on both the dark energy
equation of state and the mass selection function required to extract it.  
The likelihood of the number counts
follows a Poisson distribution whose mean fluctuates with the
large-scale structure of the universe.  
We develop a joint likelihood method that accounts for these distributions.
Maximization of the likelihood over a theoretical model that includes
both the cosmology and the observable-mass relations allows for
a 
joint extraction of dark energy and cluster structural parameters.
\end{abstract}
\maketitle

\section{Introduction}

Upcoming high redshift cluster surveys have the potential to provide
precision constraints on the evolution of the dark energy density due
to the exponential sensitivity of the abundance of massive dark matter
halos to the growth of structure (e.g. \cite{WanSte98,HaiMohHol01}).
The wealth of information contained in even the cluster counts alone
partially offsets the fundamental problem that the masses of the
clusters are not directly observable.  Techniques that utilize this
information to relate cluster observables such as Sunyaev-Zel'dovich
(SZ) flux, X-ray temperature and surface brightness, or optical
richness and shear to cluster masses and hence the cosmology are known
in the literature as ``self-calibration'' methods
\cite{LevSchWhi02,MajMoh03,Hu03a}.

Extracting the full information from the number counts requires a joint
analysis of the mean abundance and spatial distribution.  
It is also required in that
the spatial clustering of clusters is the source of sampling errors
for the counts \cite{HuKra02}.  Likewise,
cluster power spectrum estimators must account for the non-Gaussian
distribution of their errors even in the linear regime.  This fact
distinguishes cluster likelihood methods from their galaxy
counterparts \cite{VogSza96,Tegetal98} in requiring a full treatment
of the Poisson shot noise \cite{Whi79,DodHuiJaf97,Coh05}.

Likelihood analyses of the local cluster abundance in the literature
conversely include the Poisson shot noise but omit the power spectrum
source of error and information (see e.g.~\cite{PieScoWhi02,Ikeetal01}).  This
omission is justified as they typically utilize only a handful of the very
rarest clusters where shot noise dominates and the likelihood is
Poisson.  Fisher matrix studies show that for sufficiently deep
surveys the likelihood deviates significantly from the Poisson limit
\cite{HuKra02} and that these deviations contain useful information
for self-calibration \cite{LimHu04,LimHu05}.  For example, $\sim 10^4$
high redshift clusters are expected in the
upcoming South Pole Telescope (SPT) survey \footnote{\tt
  http://spt.uchicago.edu}, allowing the sample to be divided both spatially and
in SZ flux.

In this {\it Brief Report}, we develop the likelihood techniques
required to extract this information.  We show that the joint
Poisson-Gaussian likelihood can be reduced to a closed-form expression
that is applicable to the next generation high-$z$ cluster surveys.
We begin in \S \ref{sec:likelihood} with a general description of the
likelihood function and specialize it to the case of rare high-$z$
clusters in \S \ref{sec:coarse}.   We conclude in \S
\ref{sec:discussion}.

\section{Cluster Likelihood Function}
\label{sec:likelihood}

Let us take the data to be the number of clusters $N_{i\mu}$ in bins
of some observable proxy for mass ($\mu=1, \ldots, \nm$) and pixels
delineated by angle and redshift ($i = 1,\ldots,\np$).  We will assume
that the clusters are Poisson distributed with a mean of $m_{i\mu}$.
This mean fluctuates from pixel to pixel due to the large scale
structure of the universe.  We will further assume that these pixels
are sufficiently large that the fluctuations are in the linear regime.
Denoting the ensemble means with overbars,
\begin{equation}
m_{i\mu} = \bar m_{i\mu}( 1+ b_{i\mu} \delta_i )\,,
\label{eqn:fluctuatingmean}
\end{equation}
where $\delta_i$ is
 the overdensity within the pixel and $b_{i\mu}$ is the linear bias of
the selected clusters.

The distribution of Poisson means is then given by 
a Gaussian with covariance \cite{HuKra02}
\begin{equation}
\langle (m_{i\mu} -  \bar m_{i\mu}) (m_{j\nu} -  \bar m_{j\nu}) \rangle 
= \bar m_{i\mu} \bar m_{j\nu} b_{i\mu} b_{j\nu} S_{ij} \,,
\end{equation}
where
\begin{equation}
S_{ij} = \int {d^3 k \over (2\pi)^3} W_i^* ({\bf k}) W_j ({\bf k}) P(k;z_{ij})
 \,.
\end{equation}
Here $W_i$ is the Fourier transform of the pixel window, normalized such that
$\int d^3 x W_i({\bf x}) = 1$  and $P(k;z_{ij})$ is the
linear power spectrum at the mean redshift of the pixels $z_{ij}$.   

Through $N$-body simulations, a cosmological model predicts the
comoving spatial number density of clusters as a function of their
mass $M$ (e.g. \cite{SheTor99,Jenetal01,Evretal02,HeiRicWarHab04}).
Given a selection in mass $p_{i\mu}(M)$ and the comoving volume of the
pixel $V_{i}$, the ensemble mean becomes
\begin{equation}
\bar m_{i\mu} = V_{i} \int d\ln M \, p_{i\mu}(M) {d n\over d\ln M}\,.
\end{equation}
Likewise, simulations predict the bias of clusters as a function of
their mass $b(M)$ and hence
\begin{equation}
b_{i\mu} = {V_{i} \over\bar m_{i\mu}} \int d\ln M \, p_{i\mu}(M) b(M) 
{d n\over d\ln M}\,.
\end{equation}
For a completely fixed selection $p_{i\mu}$, the cosmology
uniquely determines both the set of ensemble mean counts $\bar
m_{i\mu}$ and their distribution through $b_{i\mu}$ for each pixel
$i$.  This fact allows a measure of ``self-calibration'' of the
selection function $p_{i\mu}(M)$ through the relationships between
$b_{i\mu}$ and $\bar m_{i\mu}$ in a given pixel \cite{Hu03a,MajMoh03}.
Note that the selection function can include effects such as
instrument noise and point source contamination as well as the
observable-mass relation itself.

To exploit this information, one can maximize the likelihood of the
data $N_{i\mu}$ over a theory defined by both cosmological and
selection parameters.   Fisher matrix
techniques estimate that
recovery of the dark energy equation of state at the $\sim 10\%$ level
for next generation surveys like the SPT is possible, even in the
presence of moderate uncertainties in the selection function
\cite{MajMoh03,LimHu04,LimHu05}.

In practice, extracting this information will require an exploration
of the likelihood space.  Combining the Poisson likelihood of
drawing $N_{i\mu}$ from $m_{i\mu}$ and the Gaussian likelihood of
drawing $m_{i\mu}$ from $\bar m_{i\mu}$, we obtain \cite{LimHu04}
\begin{eqnarray}
{\cal L}({\bf N} |{\bf  \bar m}, {\bf b}, {\bf S}) &=&\left[ \prod_{i=1}^{\np} \int_{-\bmax^{-1}}^{\infty}  d \delta_i
\left(  \prod_{\mu=1}^{\nm} {m_{i\mu}^{N_{i\mu}} \over N_{i\mu}! } e^{-m_{i\mu}} \right) \right] \nonumber\\
&& \times {1 \over \sqrt{(2\pi)^\np {\rm det} {\bf S}}} e^{-{1\over 2}\vecdelta^T {\bf S}^{-1} \vecdelta} \,.
\label{eqn:fullconvolution}
\end{eqnarray}
Here and throughout we use vector and matrix notation where no
confusion will arise.  The lower limit on the integral is required to
prevent negative Poisson means.  As long as $b_{i\mu} S_{ii}^{1/2} \ll
1$, the likelihood is
insensitive to the exact value of $\bmax$ and so we set it to the
maximum bias for clusters obtainable in the model.  In more 
marginal cases, the missing probability from $-1 \le \delta \le -1/\bmax$
can be restored to the $N=0$ likelihood  (see \S \ref{sec:coarse} for further discussion).
For $S_{ii}^{1/2} \simgt 1$, the density field itself is not Gaussian.

Since $m_{i\mu}$ depends on $\delta_i$ through
Eqn.~(\ref{eqn:fluctuatingmean}), the full likelihood is a convolution
of the Poisson and Gaussian distributions.  To evaluate this
convolution efficiently, let us rewrite it as
\begin{eqnarray}
{\cal L}({\bf N} |{\bf  \bar m}, {\bf b}, {\bf S})  &=&\Big[  
\prod_{i=1}^{\np} 
 \prod_{\mu=1}^{\nm} {(-1)^{N_{i\mu}} \over N_{i\mu}! } 
{\lim_{\lambda_{i\mu}\rightarrow 1}} 
\left( {\partial \over \partial \lambda_{i\mu} }\right)^{N_{i\mu}} \nonumber \\
&& \times  e^{-\lambda_{i\mu}
 \bar m_{i\mu} }\Big]
F(\bar {\bf m},{\bf b},{\bf S},{\veclambda}) \,,
\end{eqnarray}
where the generating function
\begin{eqnarray}
 F(\bar {\bf m},{\bf b},{\bf S},{\veclambda})& =&  \int_{-\bmax^{-1}}^{\infty}  d^{\np} \delta\, e^{-\sum_{i \mu} \lambda_{i\mu}
 \bar m_{i\mu} b_{i\mu} \delta_i}  \nonumber\\
 &&\times 
 {1 \over \sqrt{(2\pi)^\np {\rm det} {\bf S}}} e^{-{1\over 2}\vecdelta^T {\bf S}^{-1} \vecdelta} \,.
\end{eqnarray}
This integral may be transformed into one over a Gaussian by
completion of squares through the diagonalization of the pixel
covariance matrix ${\bf S} = {\bf A} {\vecSigma} {\bf A}^T$ where
${\vecSigma} = {\rm diag}( \sigma_k^2 )$ and ${\bf A}$ is an
orthonormal matrix of eigenvectors.  If the covariance matrix is
already nearly diagonal, the integration range will likewise transform
to a semi-infinite interval such that
\begin{eqnarray}
F(\bar {\bf m},{\bf b},{\bf S},{\veclambda})& =&
 \prod_{k=1}^{\np} e^{ {1\over 2} R_k^2 \sigma_k^2} 
[1- {1\over 2}{\rm Erfc}(y_k /\sqrt{2})] \,, 
\label{eqn:completesquares}
\end{eqnarray}
where
\begin{eqnarray}
R_k &=& \sum_{i\mu} \lambda_{i\mu} \bar m_{i\mu} b_{i\mu} A_{i k} \,, \nonumber\\
y_{k} &=& \sum_i {1 \over \bmax \sigma_k} A_{i k} - R_k \sigma_k \,.
\end{eqnarray} 
The Erfc term appears because of the boundary at $-b_{\rm max}^{-1}$ and must
be included for cases where the Poisson and Gaussian variance are comparable even
if $\bmax \sigma_k \ll 1$ as required by Eqn.~(\ref{eqn:fullconvolution}).

\begin{figure}
\epsfig{file=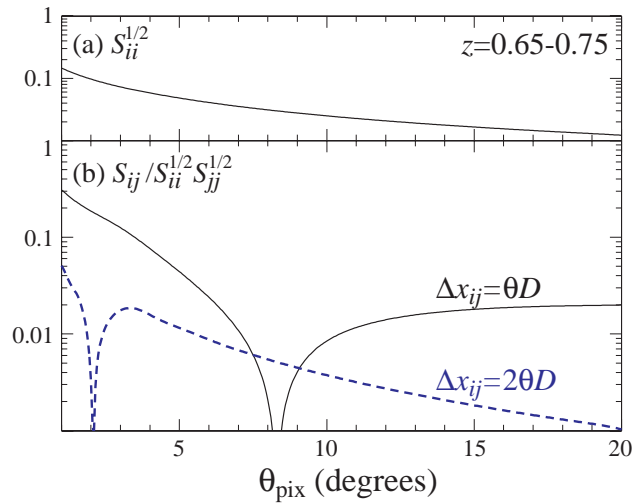, width=3.25in}
\caption{\footnotesize Covariance matrix as a function of pixel size
  $\theta_{\rm pix}$ at $z=0.7$, $\Delta z=0.1$: (a) rms density fluctuation
  per pixel $S_{ii}^{1/2}$; (b) correlation
  $S_{ij}/(S_{ii}S_{jj})^{1/2}$ for neighboring pixels.  For
  $\theta_{\rm pix} \simgt 1^\circ$ the covariance matrix is nearly
  diagonal with negligible correlations beyond the nearest neighbor
  pixel.  }
\label{fig:covariance}
\end{figure}

\section{Coarse Pixel Limit}
\label{sec:coarse}

Clusters at high redshift are extremely rare objects and hence coarse
pixelization is required to have a finite expectation value $\bar
m_{i\mu}$ per pixel.  The likelihood function simplifies considerably
in this limit.

For definiteness, let us assume that a survey like the SPT counts
clusters above an observable threshold that corresponds to a mean mass
of $M_{\rm th}=10^{14.2} h^{-1} M_\odot$ with a $\sigma_{\ln M}=0.25$
Gaussian scatter in $\ln M$ (see \cite{LimHu04} for calculational
details).  We will take as the fiducial cosmology a flat $\Lambda$CDM
model with $\Omega_{\rm \Lambda}=0.73$, baryon density $\Omega_bh^2
=0.024$, matter density $\Omega_m h^2 =0.14$ and a scale invariant
initial curvature spectrum with amplitude $\delta_\zeta =5.07\times
10^{-5}$ corresponding to $\sigma_8=0.91$.  Near the median of the
redshift distribution at $z=0.65-0.75$, the angular number density of
these clusters is $0.86$ deg.$^{-2}$ in this fiducial cosmology and
they have a linear bias of $b\approx 4$.  To have an expectation value
of at least 1 cluster per pixel, the pixels must be at least $1^\circ
\times 1^\circ$ in size.

The transverse size of the pixel is $L_{\rm pix}=D \theta_{\rm pix}
\approx 31 h^{-1}$ Mpc ($\theta_{\rm pix}/1^\circ$) where $D$ is the
comoving angular diameter distance to redshift $z\approx 0.7$.  Since
a pixel of several degrees on the side receives its fluctuations from
$k$ modes near the peak of the power spectrum, the spectrum is nearly
white and the pixel covariance matrix is nearly diagonal.

To quantify this expectation, let us take square pixels on the sky of
angular extent $\theta_{\rm pix}$ with a depth $\Delta z$ centered
around a redshift $z$.  The covariance in the radial direction is
negligible for $\Delta z =0.1$ and so we calculate only the angular
pixel covariance at each redshift slice.  The Fourier window function
is then
\begin{equation}
|W({\bf k})| = j_0(k_x L_{\rm pix} /2) j_0(k_y L_{\rm pix} /2) j_0(k_z \Delta z/2H) \,,
\end{equation}
where $H$ is the Hubble parameter.
The pixel covariance matrix becomes
\begin{equation}
S_{ij} = \int {d^3 k \over (2\pi)^3} |W({\bf k})|^2 \cos(k_x \Delta x_{ij})
\cos(k_y \Delta y_{ij})
P(k;z)
\end{equation} 
where 
$\Delta x_{ij} = L_{\rm pix} n_{xij}$ and $n_{xij}$ is the number of pixels separating
$i$ and $j$ in the $x$ direction and likewise for 
$\Delta y_{ij}$.

In Fig.~\ref{fig:covariance}, we plot the correlation of the nearest 2
neighboring pixels.  Even for $\theta_{\rm pix}=1^\circ$, the
covariance matrix beyond the nearest neighbor is negligible.
Therefore $S_{ij}$ is extremely sparse and the diagonalization can be
efficiently implemented.  In fact, for $\theta_{\rm pix} > 3^\circ$ a
strictly diagonal covariance matrix is a good approximation.

Now let us examine the likelihood function under this diagonal
covariance matrix approximation.  For simplicity, we first also assume
that there is only one bin in observable mass such that the selection
function is $
p_i(M) =   {1\over 2}{\rm Erfc}[\ln(M_{\rm th}/M) /\sqrt{2 \sigma_{\ln M}^2}]
$,
i.e.~a smooth function of the mass that increases from 0 to 1 across the threshold.
The likelihood then simplifies to ${\cal L}  = \prod_{i=1}^{\np} {\cal L}_i$, where
\begin{eqnarray}
{\cal L}_i &=&  {(-1)^{N_{i}} \over N_{i}! } 
{\lim_{\lambda_{i}\rightarrow 1}} 
\left( {\partial \over \partial \lambda_{i} }\right)^{N_{i}}   e^{-\lambda_{i}
 \bar m_{i}+{1\over 2}  \lambda_i^2 \bar m_i^2b_i^2 \sigma_i^2 }\label{eqn:diagonalform}\\
&& \times
\left\{ 1- {1\over 2}{\rm Erfc}\left[ {1 \over \sqrt{2}} \left({1 \over b_i \sigma_i} - \lambda_i \bar m_i b_i \sigma_i\right)
\right]\right\}\,.\nonumber
\end{eqnarray}

\begin{figure}
\epsfig{file=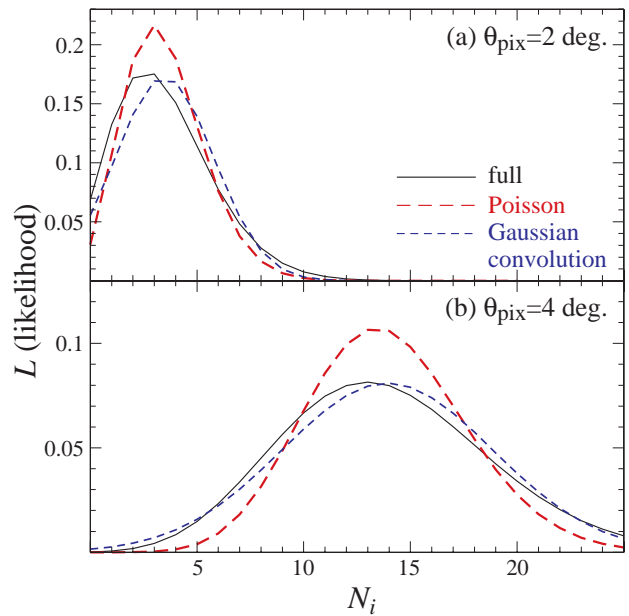, width=3.25in}
\caption{\footnotesize Likelihood as a function of counts $N_i$ above
  a mean threshold mass $M_{\rm th}=10^{14.2} h^{-1} M_\odot$ and
  $z=0.7$, $\Delta z=0.1$ for (a) $\theta_{\rm pix}=2^\circ$ and (b)
  $\theta_{\rm pix}=4^\circ$.  Shown are the full likelihood (solid)
  from Eqn.~(\ref{eqn:diagonalform}) compared with a Poisson
  likelihood (long dashed) and the $\bar m_i\gg 1$ limiting form in
  Eqn.~(\ref{eqn:convolution}) of a convolution of Gaussian shot noise
  and sample variance (dashed).}
\label{fig:likelihood}
\end{figure}

In Fig.~\ref{fig:likelihood}, we plot this likelihood as a function of
$N_i$ and compare it with the Poisson likelihood.  Here we take two
representative cases: $\theta_{\rm pix}=2^\circ$ where $\bar m_i=
3.45$, $\sigma_i= 0.097$; and $\theta_{\rm pix}=4^\circ$ where $\bar
m_i=13.9$, $\sigma_i=0.057$.  
The full likelihood is broader than the
Poisson likelihood due to the fluctuating Poisson mean.  This excess variance can be used
to self-calibrate the mass selection through the bias in a manner
that is stable to the choice of pixelization in this range \cite{LimHu04}.

This likelihood is normalized such that $\sum_{N_i} {\cal
  L}_i(N_i) =1$ for a fixed model in the limit that $b_i \sigma_i \ll
1$.  In practice, the finite $b_i \sigma_i$ will mean that there is a
finite probability of drawing a Poisson mean of $m_i = 0$ and hence no
clusters in the pixel.  This probability has been omitted in
Eqns.~(\ref{eqn:fullconvolution}) and
(\ref{eqn:diagonalform}) but may be directly restored to the $N_i=0$ likelihood.  For $\theta_{\rm
  pix}=2^\circ$, this correction is $0.005$; for $4^\circ$, this
correction is a negligible $8 \times 10^{-6}$.

For the $\theta_{\rm pix}=4^\circ$ case, $\bar m_i \gg 1$ and the
Poisson distribution of $N_i$ around $m_i$ should be nearly Gaussian.
In this limit, the full likelihood is approximately the convolution of
two Gaussians and becomes
\begin{eqnarray}
{\cal L}_i  &\rightarrow& {1 \over \sqrt{ 2\pi s_i^2 }}
\exp \left(  { (N_i - \bar m_i)^2 \over  2 s_i^2} \right)\,,
\label{eqn:convolution}
\end{eqnarray}
where $s_i^2 = \bar m_i + \bar m_i^2 b_i^2 \sigma_i^2$ is the total
variance of the convolution.  Fig.~\ref{fig:likelihood} shows the accuracy of
this approximation.

Finally, it is instructive to consider the case where the Poisson
variance is much larger than the Gaussian sample variance, i.e.  $\bar
m_{i} \gg \bar m^{2}_{i}b_{i}^2 \sigma_{i}^2$.  In this case the Erfc
boundary term is negligible and Taylor expanding the likelihood gives
a simple closed form
\begin{eqnarray}
{\cal L}_i  &\rightarrow&  { {\bar m}_{i}^{N_i}e^{-\bar m_i} 
\over N_{i}! } \left\{  1 + {1 \over 2}{ [( N_{i}-\bar m_{i} )^{2} - N_{i} ]}
{b_{i}^{2} \sigma_{i}^{2}} \right\} \,.
\end{eqnarray}
For the fiducial survey at $M_{\rm th}=10^{14.2} h^{-1} M_\odot$,
$\bar m_i b_i^2 \sigma_i^2 = 0.52$ for $\theta_{\rm pix}=2^\circ$ and
$0.76$ for $\theta_{\rm pix}=4^\circ$.  This implies that the Poisson
variance is comparable to the sample variance and so this limiting
case does not apply.  For a finer pixelization, this limit can be
reached but not before the linear theory assumption breaks down.
Nevertheless, a survey that selects rarer clusters may reach this
limit in the linear regime.  For example with $M_{\rm th}=10^{14.4}
h^{-1} M_\odot$, $\bar m_i = 4.6$ for $\theta_{\rm pix}=4^\circ$
pixels and $b_i=4.7$, and the approximation deviates by
less than $5\%$ near the peak.  It is also
useful in understanding the qualitative behavior:  the Gaussian variance of the
Poisson mean $m_i$ reduces the likelihood near the ensemble
mean $N_i = \bar m_i$ and enhance it in the tails.

Likewise, the Poisson dominated limit is useful in understanding the
effect of having multiple observable-mass bins per pixel.  For example
with 2 bins
\begin{eqnarray}
{\cal L}_i  &\rightarrow &  { {\bar m}_{i1}^{N_{i1}}e^{-\bar m_{i1}} \over N_{i1}!}   
{ {\bar m}_{i2}^{N_{i2}} e^{-\bar m_{i2}}  \over N_{i2}!} 
 \Big(  1 + {1 \over 2}\big\{  [b_{i1}( N_{i1}-\bar m_{i1} )\nonumber\\ && + 
 b_{i2}( N_{i2}-\bar m_{i2} )]^2  
 - b_{i1}^2 N_{i1}
  - b_{i2}^2 N_{i2} \big\} \sigma_i^2 
 \Big)\,.
\end{eqnarray}
The enhancement of the tails of the distribution occurs most strongly
for joint fluctuations of $N_{i1}$ and $N_{i2}$ as would be expected
from the joint fluctuations of their Poisson means $m_{i1}$ and
$m_{i2}$.

The two cases of a nearly diagonal pixel covariance and a
Poisson dominated likelihood considered here are the ones of practical
importance for high redshift clusters.

\section{Discussion}
\label{sec:discussion}

We have presented a closed-form expression for the full likelihood 
function for cluster number count surveys including
both the Poisson shot noise and Gaussian sample variance of the Poisson means
from the large-scale structure of the universe.  We also allow for multiple bins
in the observable mass per spatial pixel.

This treatment is especially useful for high redshift cluster surveys.
Here the spatial covariance is nearly diagonal for pixels that are
sufficiently large to have an expectation value of more than one
cluster per pixel.  A sparse covariance matrix allows the likelihood
to be evaluated efficiently.  Maximization of the likelihood over
models for both the cosmology and the observable-mass relations will
allow future cluster surveys to jointly determine dark energy and
cluster structural parameters.

{Acknowledgments:} We thank T. Crawford, G. Holder, and M. White 
for useful discussions. J.D.C.
was supported in part by NSF AST-0205935.
W.H. was supported by the DOE, the Packard Foundation, and the KICP
under NSF PHY-0114422.

\bibliography{HuCoh06}

\end{document}